# Low Starting Electron Beam Current in Degenerate Band Edge Oscillators

Mohamed A. K. Othman, *Student Member, IEEE*, Mehdi Veysi, Alexander Figotin, and Filippo Capolino, *Senior Member, IEEE*

*Abstract*—We propose a new principle of operation in vacuum electron-beam-based oscillators that leads to a low beam current for starting oscillations. The principle is based on super synchronous operation of an electron beam interacting with four degenerate electromagnetic modes in a slow-wave structure (SWS). The four mode super synchronous regime is associated with a very special degeneracy condition in the dispersion diagram of a cold periodic SWS called degenerate band edge (DBE). This regime features a giant group delay in the finite-length SWS and low starting-oscillation beam current. The starting beam current is at least an order of magnitude smaller compared to a conventional backward wave oscillator (BWO) of the same length. As a representative example we consider a SWS conceived by a periodically-loaded metallic waveguide supporting a DBE, and investigate starting-oscillation conditions using Pierce theory generalized to coupled transmission lines (CTL). The proposed super synchronism regime can be straightforwardly adapted to waveguide geometries others than the periodically-loaded waveguide considered here since DBE is a general property that can be realized in a variety of structures.

## I. INTRODUCTION

EFFICIENCY enhancement in vacuum high power microwave oscillators is one of the most important aspects toward achieving superior performance of microwave sources for a variety of applications [1], [2]. Slow-wave structure (SWS) modal design is critical for enhancing the interaction between an electron beam and a synchronous mode associated with the SWS.

In this paper we propose a new oscillation regime based on the synchronization between an electron beam and four electromagnetic modes that exhibit a very special degeneracy, and refer to it as *super synchronization*. This four mode super synchronization regime leads to lowering the starting-oscillation beam current, by at least an order of magnitude compared to conventional backward wave oscillator (BWO) of the same length and impedance. Accordingly, it promises potential toward enhancing the efficiency of high power oscillators. The four mode super synchronization is based on

This work was supported by AFOSR MURI Grant FA9550-12-1-0489 administered through the University of New Mexico, and by AFOSR Grant FA9550-15-1-0280.
M. A. K. Othman, M. Veysi, F. Capolino are with the Department of Electrical Engineering and Computer Science, University of California, Irvine, CA 92697 USA. (e-mail: mothman@uci.edu, mveysi@uci.edu, f.capolino@uci.edu).
A. Figotin, is with the Department of Mathematics, University of California, Irvine, CA 92697 USA. (e-mail: afigotin@uci.edu).

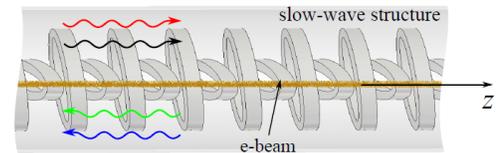
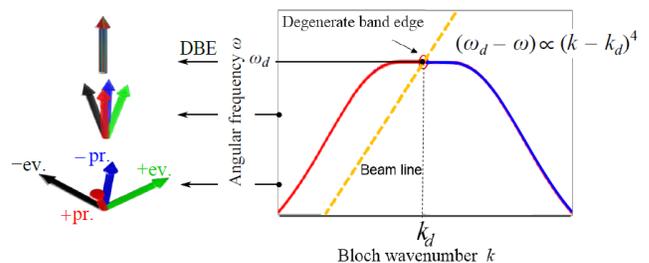

Fig. 1. An example of a slow-wave structure with four degenerate modes, and the corresponding dispersion diagram of the two propagating modes near the DBE. Four Bloch eigenmodes are shown as vectors: two of them represent propagating modes, denoted by $\pm$pr, and another two represent evanescent, denoted by $\pm$ev. The vectors coalesce at the degenerate band edge (DBE). Super synchronization occurs when the phase velocity of all waves (SWS modes and electron beam charge waves) are matched.

the degenerate band edge (DBE), which is a special point in the dispersion diagram of periodic waveguides [3]–[5], where four slow-wave degenerate modes are phase synchronized with the electron beam. This condition can be qualitatively described by having an electron beam with electrons' average velocity $u_0$ selected so that $u_0 \approx \omega_d d/\pi$, where $d$ is the period of the SWS and $\omega_d$ is the angular frequency at which the DBE in the SWS's dispersion diagram occurs. A schematic of a super synchronous oscillator made of an SWS supporting a DBE is shown in Fig. 1 with the corresponding dispersion diagram of the cold structure and the beam line. A DBE condition results in significant reduction of the group velocity of waves [3], [6] and giant increase in the local density of states [7]. These features guarantee a strong amplification in DBE-based SWS even when the electron beam is weak. The method used in this investigation is based on the Pierce model generalized to the interaction of an electron beam with coupled transmission lines (CTL), shown in [13], [14]. In [8], [9] we have introduced the concept of super synchronization leading to giant amplification in DBE-based TWTs. In particular, [8] was devoted to the study of Bloch modes in infinitely long SWS following the generalized Pierce model, and in [9] we have shown giant gain in amplifiers based on SWS with finite length. Referring entirely to the formalism in [8], [9], not repeated here for brevity, here we demonstrate a novel application of DBE-based cavities in designing low-starting current oscillators.

In particular, we demonstrate the possibility of greatly decreasing the starting-oscillation beam current using the unique DBE properties, compared to conventional oscillators. Note that the proposed principle of operation is different from the conventional absolute instability in backward wave oscillators [10]–[12]. This principle also differs from traveling wave tube (TWT) operation close to the edge of the Brillouin zone at a regular band edge (RBE) [13]–[16]. Our proposed principle advances a DBE oscillator (DBEO) that relies on synchronization of four Floquet harmonics of four degenerate Bloch modes to the electron beam.

The rest of the paper is organized as follows. We first elaborate on a representative example of a circular waveguide SWS that is periodically-loaded with elliptical metallic rings to support a DBE, similar to those demonstrated in [5]. We demonstrate the field enhancement properties as well as the giant group delay for such a SWS. Subsequently, we apply a transfer matrix method developed in [8], [9], [17], [18], to a CTL system coupled to a charge wave and examine the starting-oscillation conditions by analyzing the natural frequencies of the structure. Throughout this paper we implicitly assume a time-harmonic signal varying as $e^{j\omega t}$.

II. A LOADED CIRCULAR WAVEGUIDE WITH DBE RESONANCE

We have demonstrated in [5] that the DBE can be achieved in periodically loaded metallic circular waveguides. Here we show the DBE slow-wave resonance behavior of the "cold" periodic waveguide with finite length $L$, accounting also for losses in the metal. In addition, we focus on the group delay in such a waveguide with finite length and show that it acquires a very large value at the DBE frequency. The unconventional four mode interaction leads to a large group delay and a large field enhancement in the cold SWS (implying giant Q factors), which then serve as basis to establish a low starting-oscillation beam current operation for the SWS-electron beam interactive system. In order to investigate the proposed four mode synchronization scheme and oscillations, we develop a CTL model for such a periodic waveguide (or other optimized SWS with DBE) interacting with an electron beam, which is formulated using a generalized Pierce model [8], [17] in Sec. III.

We would like to point out that though we focus here on a specific geometry and frequency, DBE can be achieved for other analogous periodically-loaded structures currently used in TWT technology and also readily scaled to desired frequencies. The DBE resonance and its consequences including lowering starting current in analogous structures would be similar to the one investigated here.

*A. Full-wave simulation for a circular periodically loaded waveguide and DBE existence*

In Fig. 2(a) we demonstrate an example of a cutoff circular perfect electric conductor (PEC) waveguide, periodically loaded with PEC elliptical rings, that supports a DBE (losses will be considered later on in the paper). The two successive rings have an angular misalignment $\varphi$ between their major axes in the *x-y* plane, as shown in Fig. 2. Considering the design parameters of the waveguide reported in [5] (see also Appendix A), the dispersion relation between the Bloch wavenumber of the periodic structure and the angular frequency, $k-\omega$ diagram, of the four lowest order guided modes along the *z*-direction in the "cold" periodic structure is reported in Fig. 2(b), where for "cold" we mean absence of electron beam. The modes are calculated using a full-wave method (using the finite element method solver implemented in CST Microwave Studio). The dispersion relation of the four modes near $\omega_d$ is asymptotically equivalent to $(\omega_d - \omega) \approx h(k-k_d)^4$, where $h$ is a constant that depends on the geometry, that describes a very flat dispersion diagram at the edge of the Brillouin zone $k = \pi/d$. Considering the parameters in Appendix A, one obtains $h = 1.76\times10^4$ m$^4$s$^{-1}$ when the misalignment is set to be $\varphi = \varphi_{\mathrm{DBE}} = 68.8°$. It means that four fundamental modes exist near $\omega_d$: two propagating and two evanescent (with real and complex wavenumbers, respectively) and all four modes coalesce at $\omega_d$, as depicted in Fig. 1 where arrows schematically represent eigenvectors. The waveguide is designed to operate in the S-band, and the DBE frequency is set at 1.741 GHz, that is below the cutoff of the circular hollow waveguide. The resulting DBE complex TM$^z$-like mode at $\omega_d$ and $k_d$ has a strong axial electric field component (not reported here for brevity, see [5] for details). The structure supports a DBE since the rings interacts differently with two orthogonal polarizations, that are continuously coupled along the *z*-direction by changing cross section (rotating rings), such that DBE can be manifested at a specific design parameter $\varphi_{\mathrm{DBE}}$ [5] that denotes a specific ring orientations. Note that all the dispersion diagrams in this paper are calculated for lossless "cold" structures, for simplicity of visualization, whereas metallic losses are incorporated in all the following cases of transfer function, group delay, field distribution, pole location, and starting current calculations.

For a finite number of cells $N$ (taken here as $N$= 16), the termination and excitation for this particular case in Fig. 2 have been applied by utilizing 50 Ω coaxial probes, as an example. The circular waveguide is terminated at both ends by metallic walls with through holes for the probes, as seen in Fig. 2(a). The signal can be launched by a probe, denoted as port 1 in Fig. 2, into the waveguide. In the following, we consider the full-wave simulations results for a waveguide loaded by copper rings, thus taking into account metal losses (see Appendix B). We are interested in the case where the excited fields inside the waveguide create a standing wave because of the Fabry-Pérot cavity (FPC) resonance whereby constructive interference of two propagating modes with opposite group velocities leads to a sharp transmission resonance [3], [7]. Here we are examining the Fabry-Pérot resonance closest to the DBE frequency, characterized by a peak in the transmission coefficient $|S_{21}|$, close to the DBE radian frequency $\omega_d$, and approximated by [3], [6]

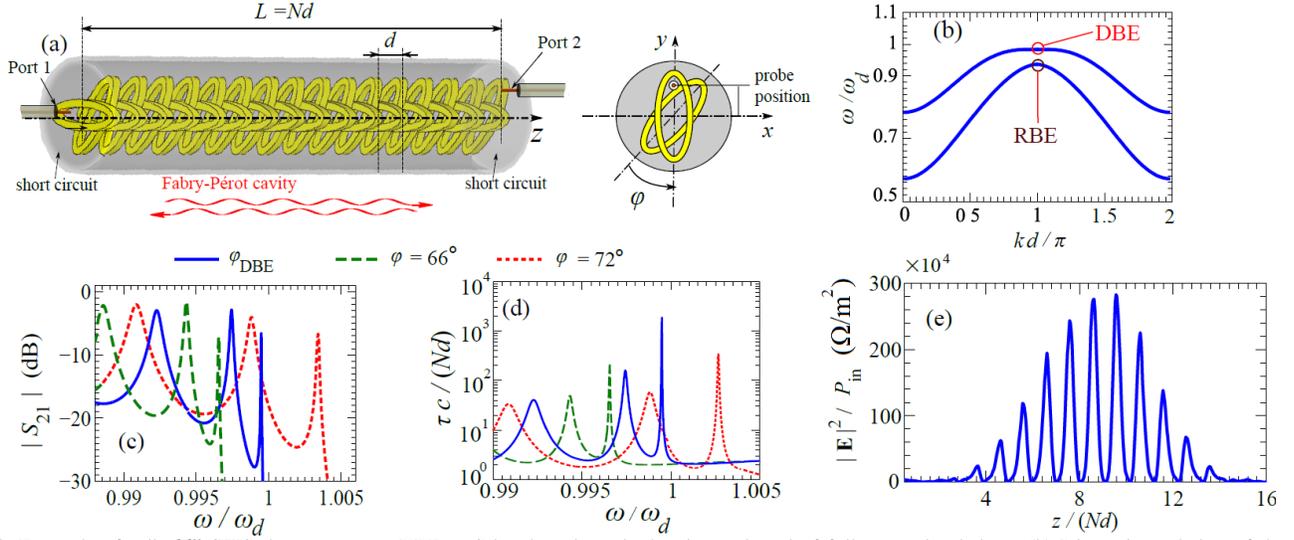

Fig. 2. Example of a "cold" SWS that supports a DBE at S-band, and results herein are based of full-wave simulations. (b) Dispersion relation of the cold periodic lossless SWS in (a). (c) Transmission coefficient $|S_{21}|$ of $N = 16$ unit cell waveguide cavity including metal losses for three misalignment angles near the DBE frequency. (d) Group delay of the $N = 16$ unit cell structure calculated evaluated as $\tau(\omega) = -\partial_\omega \angle S_{21}(\omega)$. (e) Normalized total electric field intensity on the $z$-axis of the finite length resonator.

$$\omega_{r,d} \approx \omega_d - h\left(\frac{\pi}{Nd}\right)^4. \quad (1)$$

Note that this peak tends to coincide with the DBE condition at $\omega_d$ for large number of unit cells $N$. The alignment of the Fabry-Pérot resonance with the DBE condition leads to a huge mismatch of Bloch waves at the cavity ends, since the DBE implies a complex Bloch impedance that is very hard to match [8]. Because of the DBE degeneracy condition, a giant enhancement in the energy stored in the cavity, compared to the energy escaping at the two ends of the structure, is observed, and that leads to a huge enhancement in the $Q$-factor [7]. To quantify these features, we consider the magnitude of the scattering parameter $S_{21}$ and show it in Fig. 2(d) where one can see a sharp transmission peak for the optimum design $\varphi = \varphi_{\text{DBE}}$ near the DBE frequency, as it matches the frequency of the Fabry-Pérot resonator. By changing the rings misalignment angle, considering the two cases with $\varphi = 66°$ and $\varphi = 72°$, we can readily detect the detuning of the last transmission peak frequency and its spectral width, and that is due to the tuning of the band edge dispersion flatness [9] at $k = \pi/d$. Note that the magnitude of $|S_{21}|$ at the closest peak to $\omega_d$ is less than 1 ($|S_{21}| \sim -8$ dB) due to metal losses, however high $Q$-factor is still maintained. For that reason, we focus on a rather important characteristic related to the $Q$-factor of the structure, which is the enhancement of the group delay of the *cold* structure (the group delay is also called Wigner time) classically defined as $\tau(\omega) = -\partial_\omega \angle S_{21}(\omega)$ where $\angle$ denotes the phase of the transfer function and $\partial_\omega \equiv \partial/\partial \omega$ denotes the partial derivative with respect to angular frequency. The group delay is linearly proportional to the $Q$-factor of the cavity [7]. Large group delays suggests small threshold beam currents for starting-oscillations.

In Fig. 2(d), we show the normalized group delay $\tau c/L$, where $c$ is the speed of light, for a circular waveguide with finite length $L$ and with ring tilt angle equal to $\varphi = \varphi_{\text{DBE}} = 68.8°$ as well as the two cases with $\varphi = 66°$ and $\varphi = 72°$. The normalization factor $L/c$ is the time light would take to travel the same distance $L = Nd$ in vacuum. We report that the maximum group delay occurs for the DBE case. In addition, for $\varphi = 72°$ the delay $\tau$ is 25% lower than that at the DBE condition, however it is still larger than the delay for the case with $\varphi = 66°$ because the former has a flatter dispersion near the band edge [9]. Therefore, detuning the misalignment angle from the optimal operating point corresponding to $\varphi_{\text{DBE}}$ by $\pm 4\%$ leads to reducing the group delay by ~25%. This shows how fabrication tolerances may detune a device; and also how the misalignment can be controlled for short pulsed generation as well as in pulse compression [19], in the case when a fine control of parameters is allowed. Finally, in Fig. 2(e) we show the enhancement of the total electric field intensity inside the cavity along the $z$-axis normalized to the input power. The enhancement factor is defined as $|\mathbf{E}(z)|^2/P_{in}$ where $\mathbf{E}(z)$ is the total electric field along the $z$-axis of the waveguide, and $P_{in}$ is the total input time-average power from port 1. The field is rapidly growing inside the FPC as shown in Fig. 1(e), and it reaches a maximum in proximity of the cavity center. The giant field intensity enhancement of ~$2.8 \times 10^6$ $\Omega/\text{m}^2$, despite terminations of 50 ohms coaxial cable at both ends, is a consequence of the strong excitation of both propagating and evanescent modes near the ends of the cavity as originally explained in [3], [20] for an analogous structure with DBE. These preliminary results suggest the use of the proposed example of an FPC with DBE in designing oscillators.

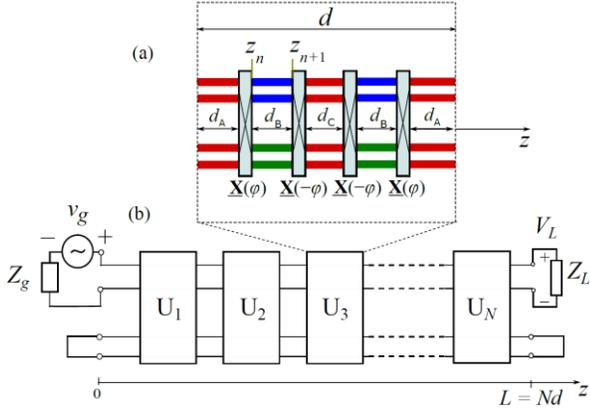

Fig. 3. (a) Coupled transmission line (CTL) model for the waveguide unit cell example in Fig. 1(a). (b) CTL model for the finite length cavity with $N$ cascaded unit cells (denoted by U), with voltage source and impedance terminations.

In the following we briefly describe the equivalent CTL model that equivalently describes fields in a periodically loaded waveguide which develops a DBE. We focus here on the important aspects of the four mode synchronization in finite length structures as well as the necessary formulation for calculating the starting oscillation conditions.

*B. Equivalent CTL model that supports a DBE*

We describe the amplitudes of the transverse electric and magnetic fields in the waveguide with the two-dimensional vectors $\mathbf{V}(z) = [V_1(z) \; V_2(z)]^T$ and $\mathbf{I}(z) = [I_1(z) \; I_2(z)]^T$, respectively, where the $T$ superscript means "transpose". We recall the equivalence of such periodic waveguides and transmission lines as formulated in [21]. We are interested in a CTL that supports two independent modes (in each direction) because it is the minimum mode count for a structure to develop a DBE [5]. That is a degeneracy of these two modes with other two with opposite sign wavenumbers, due to the $\pm z$ symmetry, at the edge of a Brillouin zone, i.e., at $k = \pi/d$. Accordingly, we adopt a CTL model shown in Fig. 3, that emulates the DBE in the waveguide, by resorting to a simple transfer matrix procedure already explained in [8], [9], [19]; and also including cut off conditions (see appendix A, and Eq. (4) in [9]). We find that, as an example, the periodic CTL in Fig. 3, whose unit cell consists of five segments of CTLs resembling the unit cell of the waveguide in Fig. 2(a), is able to support a DBE at the same frequency of the waveguide in Fig. 2(a), when using the CTL parameters provided in Appendix A. In this example, a mode mixing matrix described in [8], [9] is located at the interface between two contiguous CTL segments. This matrix, to be revisited later, intermixes modes and ensures that DBE is manifested at a certain design value of the parameter $\varphi$ (see Eq. (2) in [9]). The DBE design of the "cold" lossless CTL model is set to have $\varphi = \varphi_{DBE} = 68.8$ degrees, as to mimic the ring misalignment feature of the waveguide demonstrated earlier from full-wave simulations in Fig. 2. The "cold" lossless dispersion relation near the DBE frequency using the CTL parameters in Appendix A is plotted in Fig. 4 (a), for the three misalignment angles considered in the waveguide in Fig. 2(a). Note the resemblance of results in Fig. 4(a) with the full-wave results for the waveguide Fig. 2(b), near $k = \pi/d$, thanks to a proper choice of CTL parameters. We readily observe the detuning of both the band edge frequency and the flatness of the dispersion curve upon tuning the parameter $\varphi$ [9].

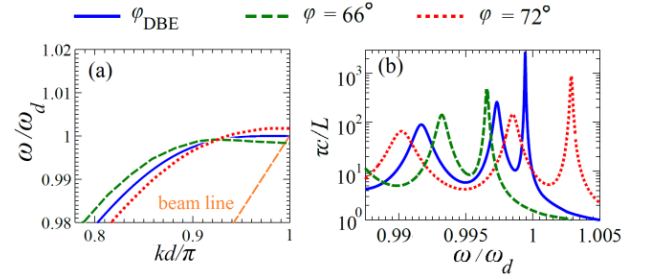

Fig. 4. (a) Detuning the "cold" lossless CTL dispersion relation near the DBE condition by changing the misalignment angle. (b) Detuning the lossy CTL resonator group delay by the same.

To further explore the analogy between the waveguide in Fig. 2(a) and the CTL in Fig. 3(a), we consider a finite length cold CTL structure composed of $N$ unit cells of the illustrated in Fig. 3(a), and calculate the CTL cold voltage transfer function defined as

$$T_F(\omega) = V_L(\omega)/v_g, \qquad (2)$$

with $V_L(\omega)$ being the load voltage at the termination impedance $Z_L$ caused by the generator voltage $v_g$. The transfer function can be derived by the transfer matrix formulation in Appendix B. In Fig. 4(b) we show the detuning of the group delay of the CLT, including losses, here calculated as $-\partial_\omega \angle T_F(\omega)$, where $\angle T_F$ denotes the phase of $T_F$ (refer to Appendix B for estimation of losses). The group delays in Fig. 4(b) are in agreement with the effect of the misalignment angle as shown in Fig. 2(d) for the waveguide, on both the magnitude and the frequency shift of the peaks of the group delay. This provides for an evident agreement between the full-wave results and the equivalent CTL model in terms of the cavity characteristics. Such agreement is a clear indication the CTL model is an effective tool that can be adapted to study SWSs with eigenmode degeneracies. Accordingly, electron beam interaction with degenerate eigenmode is formulated using the generalized Pierce model developed in [8], [19], and elaborated in the following.

### III. CTL Interacting With an Electron Beam

*A. CTL-beam Interaction model*

The interaction between the SWS modes and an electron beam is investigated via a CTL approach, extending the single mode linear interaction theory developed by Pierce and contemporaries [22]–[25] to multimode interaction, as was done in [8], [17], [18], [17]. Accordingly the beam is described by modulation of the *charge wave* current $I_b$ and beam velocity $u_b$ of the total electron beam current $I_0 + I_b$ and velocity $u_0 + u_b$, respectively. It is convenient to define an equivalent kinetic voltage modulation $V_b = u_0 u_b / \eta$ [8], [17], [26], with the same time harmonic frequency as the electromagnetic field in the SWS because the charge wave velocity modulation will have the same frequency of the electromagnetic field. Here $\eta = e/m = 1.758 \times 10^{11}$ C/kg is the electron charge-to-mass ratio, and $-e$ and $m$ refer to the charge and mass of the electron, respectively. Such charge wave describes the bunching and debunching of electrons which causes energy exchange between the beam and the modes in the SWS. The unmodulated electron beam has an average (d.c.) current $I_0$ and a d.c. (time-average) electron's kinetic energy $eV_0$, where $V_0$ is the beam voltage [27], [28] that is related to the average non-relativistic electron velocity $u_0$ by $V_0 = u_0^2/(2\eta)$. When coupled to EM fields, the total beam current is $I_0 + I_b$ with $|I_b| \ll I_0$ and the total equivalent beam voltage is $V_0 + V_b$ with $|V_b| \ll V_0$, based on small-signal considerations.

Here, as in [8], [9], [17] we adopt a space varying *state vector*

$$\Psi(z) = \begin{bmatrix} \mathbf{V}^T(z) & \mathbf{I}^T(z) & V_b(z) & I_b(z) \end{bmatrix}^T, \quad (3)$$

composed only of the field quantities that vary along the $z$-direction, which are the transmission line voltage and current vectors, as well as charge wave quantities $I_b$ and $V_b$. The system of linear equations are developed in [8], [9], based on small beam modulations in accordance to the Pierce model. The evolution of the state vector is conveniently represented by a first order partial differential transport equation form as $\partial_z \Psi(z) = -j\underline{\mathbf{M}}(z)\Psi(z)$, where $\underline{\mathbf{M}}(z)$ is the 6×6 system matrix that describes all the $z$-dependent CTL, electron beam and space-charge parameters, including coupling effects and losses [8], [9], [17]. Accordingly, the relation of the state vector at the boundaries $z_n$ and $z_{n+1}$ of each segment in Fig. 3(a) is given by

$$\Psi(z_{n+1}) = \underline{\mathbf{T}}(z_{n+1}, z_n)\Psi(z_n), \quad (4)$$

where here we assume that $z_{n+1} > z_n$. Within a homogenous segment (where the matrix $\underline{\mathbf{M}}$ is $z$-invariant) the 6×6 transfer matrix is then constructed as $\underline{\mathbf{T}}(z, z_0) = \exp\left[-j(z - z_0)\underline{\mathbf{M}}\right]$.

Note that for a unit cell made of cascaded segments the transfer matrix is constructed using the group property such that the transfer matrix for the unit cell, $\underline{\mathbf{T}}_U$, is given by the product of segments' transfer matrices. The state vector evolves across the unit cell as

$$\Psi(z+d) = \underline{\mathbf{T}}_U \Psi(z), \quad (5)$$

where $d$ is the SWS period (see Figs. 1, 2 and 3).

The waveguide in Fig. 2 supports a DBE due to the angular misalignment of the rings' axes. Consequently the modes in the loaded waveguide are mixed from one segment to the other. With that in mind we employ a 6×6 mode mixing matrix $\underline{\mathbf{X}}(\varphi)$ in the CTL that couples voltages and currents between contiguous waveguide segments. The charge wave is assumed not to be affected by this matrix, more details on this technique is found in [8], [9]. As a result, in the chosen CTL in Fig. 3, the unit cell's transfer matrix is found as a product of the individual waveguides segments and the mode mixing matrix $\underline{\mathbf{X}}(\pm\varphi)$. The transfer matrix of the unit cell reads

$$\underline{\mathbf{T}}_U = \underline{\mathbf{T}}_A \underline{\mathbf{X}}(\varphi) \underline{\mathbf{T}}_B \underline{\mathbf{X}}(-\varphi) \underline{\mathbf{T}}_C \underline{\mathbf{X}}(-\varphi) \underline{\mathbf{T}}_B \underline{\mathbf{X}}(\varphi) \underline{\mathbf{T}}_A, \quad (6)$$

where $\underline{\mathbf{T}}_A$ and $\underline{\mathbf{T}}_C$ are the equivalent CTL transfer matrices of the two empty waveguide segments, respectively, and $\underline{\mathbf{T}}_B$ is that of the waveguide segment loaded with rings (see Fig. 3(a) for the CTL representation of the waveguide unit cell). Note that the transfer matrix $\underline{\mathbf{T}}_B$ of waveguide segments with rings is the same, however the misalignment angle is taken into account through the matrix $\underline{\mathbf{X}}(\pm\varphi)$, and in our case $\underline{\mathbf{X}}(\pm\varphi)$ is represented in terms of rotation matrices (equations (2) and (3) in [9] or equations (12) and (13) in [8] ).

*B. Four degenerate modes super synchronous condition*

The main utility of the CTL approach summarized above (thoroughly detailed in [8], [9]) is to explore a novel unconventional oscillation scheme associated with a finite length DBE cavity (such as the structure in Fig. 2(a) or analogous ones [5]) interacting with an electron beam. As we have shown above, the CTL system depicted in Fig. 3(a) does support a DBE with four degenerate Bloch modes. We would like to point out that each of the four Bloch modes can be decomposed in infinitely many Floquet harmonics[8]. Because of the degeneracy, in the "cold" SWS structure all Fouquet harmonics have wavenumbers $k_n^m = \pi/d + F_m$, with $F_m = 2\pi m/d$, and where $m = 0, \pm 1, \pm 2, \ldots$ is the index of the Floquet harmonic. Only a slow harmonic is phase synchronized with the electron beam having average velocity $u_0$, which we refer to it with the $m = 0$ index. Hence, there are four slow Floquet harmonics with identical wavenumbers $k_d = \pi/d$ at the band edge angular frequency $\omega_d$, and the super synchronism condition can be represented by the following equality

$$u_0 \approx \frac{\omega_d}{k_d} = \frac{\omega_d d}{\pi}. \quad (7)$$

The condition (7) is the necessary preliminary criterion for the oscillation scheme proposed in this paper based on the four degenerate modes synchronization. The desired $u_0$ is found by solving (7), which is depicted graphically in Fig. 4(a), by intersecting the "cold" structure dispersion and the beam line $\omega(k) = k u_0$, at the Brillouin zone edge. The dispersion diagram of the hot structure is somewhat perturbed for small electron beam current $I_0$ [8], but it is rather significantly distorted for large beam currents implying that strictly speaking one may lose the degeneracy for large beam current, though advantageous characteristics may occur nonetheless and need to be verified numerically. Therefore, the average beam velocity $u_0$ has to be finely tuned to achieve the best synchronization as we discuss in the following. We point out once again that the DBE can be obtained also in other loaded waveguides which differ from that in Fig. 2. Moreover, different CTLs than the one in Fig. 3 can emulate DBE properties in realistic waveguides. Still, the generalized CTL Pierce model can be adopted for all these situations and the conclusions derived here based on four degenerate mode synchronization would be analogous for other SWSs.

In the case of SWS with *finite length L*, as in this paper, the FPC resonance condition occurs at a frequency $\omega_{r,d}$ given in (1), that tends to $\omega_d$ for large number of unit cells *N*. For super synchronization to manifest in the finite length cavity, proper adjustment of the electron's velocity is required. Strictly speaking, though being close to the DBE condition (since $\omega_{r,d} \cong \omega_d$), at $\omega_{r,d}$ there exists two propagating modes and two evanescent ones in the cold SWS. The best design of the oscillator (in terms of starting current) is achieved by finely tuning $u_0$ in the vicinity of $\omega_d d / \pi$. For simplicity, here we choose the electron's phase velocity to match that of the propagating forward mode in the +$z$-direction at $\omega_{r,d}$ of the finite length cold cavity (without seeking any optimization). The resulting value for $u_0$ is slightly larger compared to one provided by condition (1), that is the case of an infinitely-long periodic structure (see more details about super synchronism in [8], [9]). However, the other three modes play a pivotal role for we are still very close to the degeneracy condition, and they are responsible for generating the giant DBE resonance [3], [9], [20], at $\omega_{r,d} \approx \omega_d$. Therefore their contribution in enhancing the interaction of the electron beam with that synchronized mode is essential [9]. In summary, the operational frequency slightly differs from $\omega_d$, but the cold SWS is still undergoing a degeneracy, proven by the giant group delay at $\omega_{r,d} \approx \omega_d$ shown in Fig. 4(b). Hence we can conclude that the four mode super synchronization concept summarized by the simple equation (7) is still valid. Indeed, for large enough number of unit cells *N*, based on (1), all modes of the SWS with DBE almost coincide. This new synchronization scheme leads to a significantly lower oscillations threshold than that in conventional single mode oscillators as shown next.

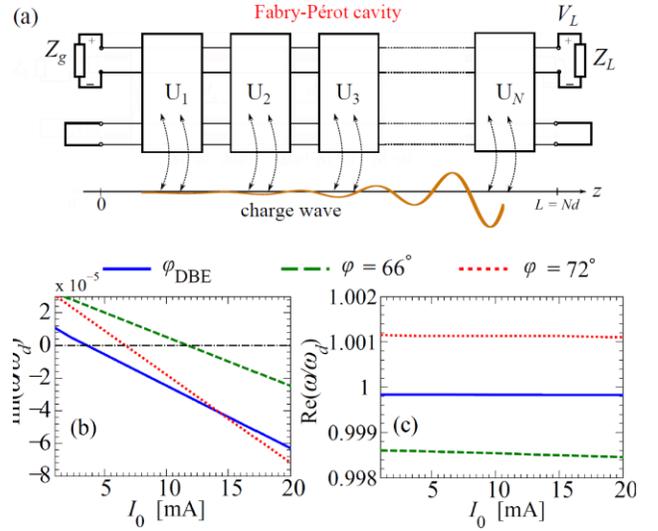

Fig. 5. (a) Schematic of the equivalent CTL-electron beam model of degenerate band edge oscillator (DBEO). (b) Real and (c) imaginary parts of the complex radian frequency pole $\omega_{\text{pole}}$ (the one closest to the DBE) of the transfer function $T_F(\omega)$ of the FPC in Fig. 3, with $N = 32$, interacting with an electron beam of beam average current $I_0$. For the optimal condition (blue curve, associated to the ideal tilt angle parameter $\varphi = \varphi_{\text{DBE}}$) the pole enters the unstable region $\text{Im}(\omega) < 0$ with the smallest beam current. Results are shown also for two other angle parameters misaligned with respect the DBE condition. Note also the insensitivity of the pole's real part (i.e., the oscillation frequency) to the beam current $I_0$.

## IV. THE DEGENERATE BAND EDGE OSCILLATOR

There are a few approaches in analyzing stability of an amplifying system and prospects of oscillations. In an infinitely long structure, the criteria for oscillations can be drawn from the dispersion diagram. Namely there can exist a gap in the real wavenumber spectrum in which frequency can only be complex with a negative imaginary part, indicating absolute instabilities [29]. The instability can also be understood from the criterion developed by Briggs [30]. We are interested in absolute instabilities in a finite length SWS, that facilitates a feedback mechanism due to mismatch at its termination by forming an FPC. Following the development of the transfer function in the preceding section and the idea of super synchronism (5), we perform a natural frequencies analysis by investigating the location of complex frequency poles of the transfer function $T_F(\omega)$ in the *complex* angular frequency $\omega$-plane as function of beam current $I_0$, (refer to Appendix C for more details), for the CTL in Fig. 4(a) under super synchronous operation. In the following cases the super synchronous condition is established by selecting a beam velocity $u_0 \cong 0.49c$ that is close to $\omega_d / k_d \cong 0.47c$, for the CTL parameters in Appendix A.

The Barkhausen stability criterion [31], [32] for starting-oscillation in feedback systems states that the positive feedback in an active device must have (i) a phase shift of an integer multiple of $2\pi$ for a roundtrip propagation of the signal inside the cavity, and (ii) the loop overall gain should be larger than unity. The former is realized through the Fabry-Pérot resonance condition in proximity of $\omega_{r,d}$ (it would be exactly at $\omega_{r,d}$ in the cold structure). The latter condition is realized by having the beam as a source of amplification with sufficient energy transfer to the electromagnetic modes to compensate for the round trip losses in the system and at the two ends of the FPC. The complex frequency poles $\omega_{\text{pole}}$ of the transfer function $T_F(\omega)$ detain the information of the resonance frequency as well as the Q-factor of the cavity. Those poles are symmetrically distributed in the complex $\omega$-plane, in pairs with $\omega_{\text{pole}\pm} = \pm \text{Re}(\omega_{\text{pole}}) + j\,\text{Im}(\omega_{\text{pole}})$. The poles are located so that $\text{Im}(\omega_{\text{pole}}) > 0$ for a stable system, whereas poles with $\text{Im}(\omega_{\text{pole}}) < 0$ dictates an unstable system. Poles determine the time-domain response for an arbitrary input under linear operation. Poles with $\text{Im}(\omega_{\text{pole}}) < 0$, revealing unstable condition, are responsible for the generation of zero-drive oscillation (in this case zero-drive means no input of electromagnetic signal in the CTL). Since the transfer function $T_F(\omega)$ has several poles, we pay special attention to the pole closest to the DBE frequency because it is the one that is closest to the real axis of the complex $\omega$ plane and causes the largest peak (highest Q factor) in Fig. 4(b). Furthermore, it has the highest sensitivity to the coupling to the charge wave, as expected from the calculation of the group delay in Figs. 2 and 3. For the sake of simplicity, we first consider a lossless FPC. The effect of metal losses on the starting current will be demonstrated at the end of this section.

In Fig. 5(b, c) we plot both imaginary and real parts of the pole nearest to the DBE resonance angular frequency $\omega_{r,d}$ (only the pole with positive real part is shown for brevity), as a function of the electron beam current $I_0$, of the transfer function (2) for a lossless CTL and with generator and load locations as in Fig. 3. The plot in Fig. 5(b) indicates that for small values of the beam current, this particular pole lies in the stable region ($\text{Im}(\omega) > 0$) whereas increasing the beam current transitions the pole into the unstable region (where $\text{Im}(\omega) < 0$), and that corresponds to triggering of oscillation. This transition of the pole manifests the onset of absolute instabilities and designates the *starting-oscillation current* or the *threshold beam current* condition, $I_{\text{st}}$. We would like to point out that the beam-SWS interaction affects mainly the poles near the DBE frequency since we have designed the SWS-beam interaction according to the super synchronous condition by selecting the beam velocity $u_0$ to be close to the phase velocity of the four modes (7) under degeneracy condition. This can be inferred by observing that other poles of the transfer function, whose real part is below the DBE resonance frequency. In other words, poles other than the one associated with the DBE resonance for which $|\text{Re}(\omega_{\text{pole}})| \approx \omega_{r,d}$, are not significantly affected by increasing the beam current in the range shown in Fig. 5 (those other poles are not reported in Fig. 5). This means that if the beam current is in a vicinity of the threshold shown in Fig. 5, then the effect of the pole close to the DBE frequency dominates over effects of all other systems poles. Moreover, by investigating changes in the misalignment parameter $\varphi$, ($\varphi = 66°$ and $\varphi = 72°$ in addition to $\varphi = \varphi_{\text{DBE}} = 68.8°$) we found that the DBE case exhibits the zero-crossing of the pole's imaginary part with the lowest beam current compared to other cases in Fig 5. Furthermore, we observed that the real part of that particular pole for $\varphi = \varphi_{\text{DBE}} = 68.8°$ is not sensitive to varying the beam current, i.e., $\partial \text{Re}(\omega_{\text{pole}})/\partial I_0 \approx 0$ for the blue curve as seen from Fig. 5(b). Interestingly the two other cases in proximity of the DBE have almost constant $\text{Re}(\omega_{\text{pole}})$ also. This analysis shows that increasing the beam current a threshold to start oscillation is encountered for the DBE case and also for the two other cases where a DBE condition is not met exactly. Accordingly, oscillation is associated with the DBE resonance mode, and the frequency of oscillation would be very close to $\omega_{r,d}$. Furthermore, in all these cases the real part of the $\omega$-pole is not strongly depended on the beam current, a condition that may lead to a purer spectral quality of the oscillation and it could be further studied using time domain techniques including saturation effects.

With the threshold condition in Fig. 5, the *starting-oscillation current* or the *threshold beam current* condition, $I_{\text{st}}$ is defined as the electron beam current such that $\text{Im}(\omega_{\text{pole}}) = 0$. We then calculate in Fig. 6 the starting current, $I_{\text{st}}$ for the lossless FPC with DBE (relative to $\varphi = \varphi_{\text{DBE}} = 68.8°$) and also for the lossless FPCs with two different misalignment parameters $\varphi$ of 66 and 72 degrees, varying as a function of the number of unit cells N (losses will be considered in Fig. 7). We have found that the optimal design with $\varphi = \varphi_{\text{DBE}}$ demonstrates a very low starting current. The other two cases with slight deviation from $\varphi_{\text{DBE}}$ also show low starting current but the starting current for the optimal design is the lowest. We also show the threshold current trend for large N, for the DBE case, by plotting the fitting curve

$$I_{st} = \frac{\alpha}{N^5}, \tag{8}$$

with $\alpha = 9 \times 10^4$ [A]. The fitting curve agrees with the threshold current values obtained by the analysis based on transfer matrix discussed previously when N is large (i.e., N > 16) which indicates that the starting current for the DBE case scales as $\propto 1/N^5$ posing a possible great advantage of utilizing DBE in high power oscillators when increasing the length. The reasons for such unconventional scaling is that Q-factor for DBE resonators scales as $Q \sim N^5$ [3], [7], [33]. For the sake of comparison we then also calculate the starting current in a standard backward wave oscillator (BWO) of the same length and with interaction impedance Z equal to the average impedance of the CTL in Fig. 3 using the formula $I_{st} = V_0 /(8ZN_\lambda^3)$ [34], [35]. Here $N_\lambda = L/\lambda_g$ is ratio between the physical length of the BWO ($L = Nd$) and the guided wavelength $\lambda_g = 2\pi/k_d$, evaluated at $\omega = \omega_d$, $V_0 = u_0^2/(2\eta)$ as in Sec. III, and the impedance $Z=51.6\ \Omega$ is calculated as in Appendix B. We observe that for structures with number of unit cells such that $N>16$ the starting current of the DBE is lower than that of a BWO. Especially, for $N = 48$ unit cell the starting current is two orders of magnitude lower than that of a standard BWO.

Next, we investigate the impact of the space charge effect that causes debunching, as well as SWS losses, on the starting condition. The space charge effect due to internal forces in the electron beam causes charge waves to oscillate with the plasma frequency $\omega_p$ [2], [23], [27] (see Appendix B). The debunching phenomenon as the space charge effect is accounted for through a plasma frequency defined as $\omega_p^2 = 2V_0 u_0/(\varepsilon_0 A)$, where $A$ is the beam cross sectional area and $\varepsilon_0$ is the vacuum permittivity [23]. The debunching effect counteracts the bunching in the electron beam and often limits the energy transfer from electron beam to electromagnetic modes in the SWS [27]. Consequently, the starting current is affected by the space charge effect. The space charge is accounted for in the evolution of the system vector $\mathbf{\Psi}(z)$ as described in Eq. (2) in [16], and in Eq. (4) in [17], and is incorporated in the transfer matrix. We investigate how the plasma frequency alters the starting-oscillation condition. This is achieved by only changing the beam area $A$, whereas all other structural parameters are kept the same. Moreover, losses affect the quality factor and the oscillator performance, thus in order to provide a meaningful assessment, the case of a lossy CTL has been investigated as well. A generic loss in the CTL is accounted for by considering a distributed series resistance in the CTL equal to $0.2\ \Omega/m$ as described in Appendix B, retrieved from fitting data obtained by full-wave simulations of the "cold" periodic ring-loaded waveguide in Fig. 2(a).

debunching (space-charge) effects. Recall that in the conventional BWO, the value of $\omega_p/\omega_d$ is a measure of the space charge effect, and is related to the Pierce space charge parameter (see Appendix A). We consider a DBEO and a BWO with equal length $L = Nd$ with $N =32$. We recall that for a lossless SWS made of $N = 32$ unit cells, when space charge effects on debunching is ignored (i.e., one may assume that the beam area $A \to \infty$ or $\omega_p \to 0$) the starting current condition is that obtained in Fig. 6, leading to a DBEO starting current $I_{st} \cong 4.1\ \text{mA}$ to be compared with the BWO $I_{st} \cong 45.3\ \text{mA}$, as shown in Fig. 6. The case of a DBEO including losses and debunching induced by space charge effects is shown in Fig. 7 (solid blue curve) as a function of the normalized plasma frequency $\omega_p/\omega_d$ in the range $0.01 \le \omega_p/\omega_d \le 0.07$. This result shows that the DBEO starting current is slightly increasing when space charge effect increases.

For the BWO, the starting current accounting for space charge effects is not simply given by $I_{st} = V_0/(8ZN_\lambda^3)$ but rather evaluated by solving the Pierce-like equations for poles following the work by Kompfner and Williams [12] and by Johnson [13]. Some details are reported in Appendix A. Comparing the threshold current for DBEO and BWO in Fig. 7 we find that the DBEO leads to a threshold that is ~30 times lower than that of the BWO also when space charge effects on debunching are accounted for.

Results in Fig. 7 also show that when losses are accounted for, the starting current is larger than in absence of losses as expected, for both the DBEO and BWO. However, for the DBEO the starting current is at least half of that of the BWO, keeping the same length $L$, same impedance and same space charge parameters.

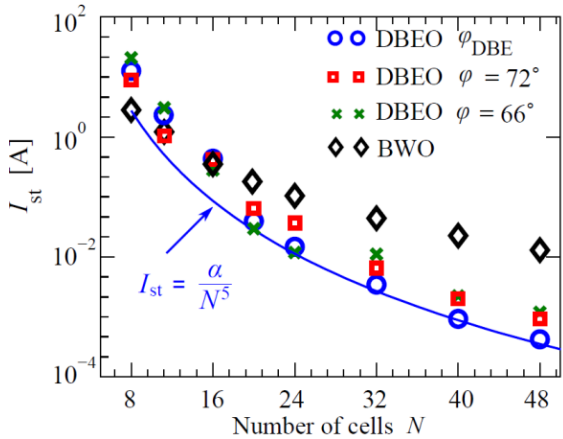

Fig. 6. Starting beam current (oscillation threshold current) for the DBEO, made of a lossless FPC with DBE, varying as a function of TL length $Nd$, under super synchronous condition (7). Losses are considered in Fig. 7. We also show starting current results for two cases with angles misaligned with respect to DBE condition whose $\varphi_{DBE} = 68.8$ degrees. For comparison the starting current for a backward wave oscillator of the same length is also plotted showing its much higher starting current.

A comparison between the proposed DBEO and a standard BWO is carried out in Fig. 7 accounting for both losses and

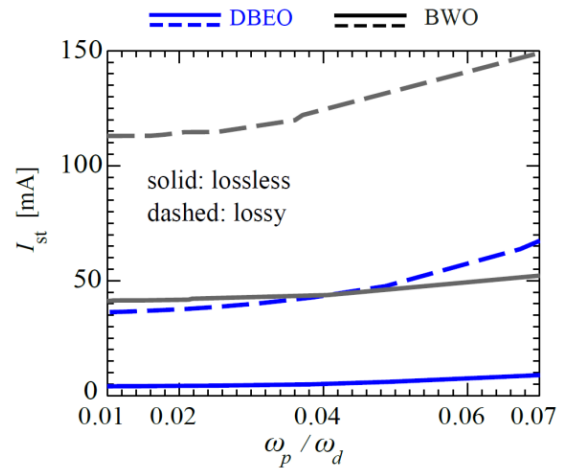

Fig. 7. Comparison between a DBE oscillator (DBEO) and BWO for both lossless case and lossy case with distributed losses of 0.2 ohm/m, varying the space charge effect represented by the normalized plasma frequency $\omega_p/\omega_d$. The oscillator length is fixed at $N = 32$.

We expect the advantages of the DBEO to be more pronounced if the SWS is carefully designed to minimize losses. We observed that the starting current increases monotonically as the normalized plasma frequency increases from 0.01 to 0.07, for both BWO and DBEO, as expected. Nevertheless, the starting current for the DBEO is much lower than that for the BWO. In summary, the starting beam current for the DBEO is maintained smaller than that of a standard BWO of the same length, taking both space charge effects and losses into account.

## V. CONCLUSION

We have demonstrated lower starting-oscillation beam current regime in slow-wave structures that exhibit a four-mode DBE in their dispersion relation, compared to the conventional single mode Pierce-type BWO. We have computed the characteristics of a finite length cold SWS with DBE using full-wave simulations as well as its equivalent CTL model. The super synchronous regime is based on four mode synchronization with an electron beam that provides for at least an order of magnitude lower starting-oscillation conditions in the DBE oscillator (that we named DBEO) than a comparable BWO, when losses and the space charge effects were taken into account. The proposed principle of four mode synchronization demonstrates a potential advantage of DBE structures for designing electron-beam-based sources at microwave and even millimeter wave frequencies.

## ACKNOWLEDGMENT

This research was supported by AFOSR MURI Grant FA9550-12-1-0489 administered through the University of New Mexico. The authors acknowledge also support by AFOSR Grant FA9550-15-1-0280. The authors would also like to thank the Computer Simulation Technology (CST) for providing CST Microwave Studio that was instrumental in this work.

## APPENDIX A: PARAMETERS USED IN NUMERICAL SIMULATIONS

The parameters of the waveguide in Fig. 2 are as follows: ring length of 15 mm, rings separation of 2.5 mm, ring thickness of 5 mm, major inner radius of 25 mm, and the aspect ratio is 2.5:1. The period is 40 mm, equal to the waveguide radius. For the finite length waveguide, the probe has a length of 22 mm and positioned at 23 mm from the center of the waveguide (i.e., in Fig. 2(a) the coaxial probe from one end is positioned on the y-axis at $y$ =23 mm, and the other probe from the other side is the same but rotated by $\varphi$). The excitation is incorporated with a coaxial TL of outer radius of 2.5 mm and inner radius of 1.2 mm and the copper conductivity is taken here as $6.3 \times 10^7$ S/m at 1.7 GHz. The equivalent CTL parameters, constitutive inductances, capacitances and cutoff capacitances per unit length for all A,B, and C segments of the CTL system (Eq. (4) in [9]), are summarized here: for TL A we assume the inductance per unit length $L = 0.33$ µH/m, capacitance per unit length $C = 2.9$ nF/m, $C_c = 80$ fFm, and $d_A = 2.5$ mm (the middle segments length is $d_A/2$), for TL B we assume $L = 3.1$ µH/m $C = 0.3$ nF/m, $C_c = 0.2$ pFm and $d_B = $ 15 mm, and for TL C we assume $L = 0.6$ nH/m, $C = 0.3$ nF/m $C_c = 0.2$ pFm and $d_C = 15$ mm. These parameters are chosen (by numerical fitting) such that the dispersion relation of the CTL resembles that of the waveguide in Fig. 2 obtained with full-wave simulation, and maintaining an average impedance $Z$ of the unit cell constituent TLs of ~51.6 $\Omega$ [9]. The average beam voltage used is in the order of $V_0 \approx 60$ kV. It is selected, as a function of the beam electrons' velocity, so that the beam is synchronized with the four mode DBE resonant peak. Note that experimental studies, as those in [36]–[38], used the above beam voltage with beam currents up to few amperes. In this paper we have used lower values of the beam current.

## APPENDIX B: SPACE CHARGE PARAMETERS AND LOSSES

The debunching phenomenon as the space charge effect is accounted through a plasma frequency defined as $\omega_p^2 = 2V_0 u_0/(\varepsilon_0 A)$, where $A$ is the electron beam cross sectional area and $\varepsilon_0$ is the vacuum permittivity [23] and it affects the beam dynamics as in Eq. (4) of [9]. We compare our results with a standard Pierce-like BWO, which has a characteristic impedance equal to the average impedance of the unit cell of the CTLs considered here, $Z = 51.6$ $\Omega$. The Pierce parameter $C_p = I_0 Z/(4V_0)$ [10], [23] is useful to assess the interaction of the modes with the electron beam in TWTs and BWOs under synchronous operation. It is also conventional to define a space charge parameter $q$ related to the Pierce parameter to investigate the effect of space charge, and $q$ is defined as $q = \left[ \omega_p/(\omega C_p) \right]^2$. In general, it is important to design TWTs or BWOs with large values of $C_p$ that give rise to better performance in terms of gain and lower space charge effects. In our simulations for the Pierce model of the BWO operating at $\omega = \omega_{r,d}$, with the impedance and beam parameters in Appendix A, we obtain $C_p = 0.02$, which is a typical value [2], [27]. Accordingly, $q$ is found to be $\approx 11.89$ when $\omega_p/\omega_d = 0.07$. This value $\omega_p/\omega_d = 0.07$ is the maximum we have used in Fig. 7 because it leads to very strong debunching due to space charge effects [23], [27].

Losses in the waveguide, considered in the CTL, have been estimated by carrying out a full-wave simulation at the DBE resonance frequency of the metallic waveguide structure assumed to be made of copper. When incorporating the finite conductivity of the copper in the finite element full-wave simulator, we calculate the total power loss by the waveguide structure. Then, by a fitting procedure, we find the equivalent per-unit-length series resistance (in series with the inductance per-unit-length) in all the TL segments of the CTL, in such a way that the CTL provides the same power loss and transmission characteristics seen in full-wave simulations. In summary, the equivalent waveguide losses are represented by a distributed CTL series resistance found to be $\cong 0.2$ $\Omega$/m. To calculate the starting current for a BWO, we use equations

(18) and (21) in Johnson [10], which accounts for the space charge effect and losses.

## APPENDIX C: TRANSFER FUNCTION AND POLES

We summarize the general formulation to obtain the transfer function of a finite length interactive system in order to calculate its poles and consequently the starting-oscillation beam current, with imposing general boundary conditions as follows. A generator voltage described by the vector $\mathbf{v}_g = \begin{bmatrix} v_{g1} & v_{g2} \end{bmatrix}^T$ with a source impedance of $\underline{\underline{\mathbf{Z}}}_g = \mathrm{diag}(Z_{g1}, Z_{g2})$ is located at $z = 0$ as in Fig. 3, i.e., the boundary condition is written as $\mathbf{V}(0) = \mathbf{v}_g - \underline{\underline{\mathbf{Z}}}_g \mathbf{I}(0)$. At a distance $L = Nd$ the CTL is terminated by a 2×2 impedance matrix $\underline{\underline{\mathbf{Z}}}_L$. We assume that the e-beam enters the interaction area at $z = 0$ without any pre-modulation, i.e., $I_b(0) = V_b(0) = 0$. Therefore we have six boundary conditions, which are sufficient to provide a unique transfer function for the coupled system. Let us define a 2×2 matrix voltage transfer function $\underline{\underline{\mathbf{T}}}_F(\omega, I_0)$ which provides the voltage at the loaded impedance $\underline{\underline{\mathbf{Z}}}_L$ subject to a generator voltage via $\mathbf{V}(L) = \underline{\underline{\mathbf{T}}}_F(\omega, I_0) \mathbf{v}_g$ as a function of complex frequency and beam current in general. The steps to find $\underline{\underline{\mathbf{T}}}_F(\omega, I_0)$ are as follows: first we recall the transfer matrix of a single unit cell. The finite length CTL is composed of $N$ unit cells, therefore we compute the transfer matrix of the $N$-cell structure namely $\mathbf{\Psi}(L) = \underline{\mathbf{T}}^N \mathbf{\Psi}(0)$, which is evaluated as $\underline{\mathbf{T}}_U^N = \underline{\mathbf{U}} \underline{\mathbf{\Lambda}}^N \underline{\mathbf{U}}^{-1}$, and may be conveniently represented by a matrix in the following block form

$$\underline{\mathbf{T}}_U^N = \begin{pmatrix} \underline{\underline{\mathbf{A}}} & \underline{\underline{\mathbf{B}}} & \underline{\underline{\boldsymbol{\alpha}}} \\ \underline{\underline{\mathbf{D}}} & \underline{\underline{\mathbf{E}}} & \underline{\underline{\boldsymbol{\beta}}} \\ \underline{\underline{\boldsymbol{\gamma}}} & \underline{\underline{\boldsymbol{\zeta}}} & \underline{\underline{\boldsymbol{\sigma}}} \end{pmatrix}. \quad (B1)$$

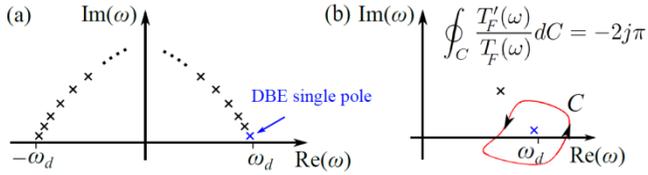

Fig. B1. (a) Typical pole distribution of an FPC with DBE in the complex angular frequency $\omega$ space, very close to the DBE frequency. (b) Application of Cauchy argument principle to verify the existence of a single pole near the DBE frequency.

We wish to write an expression for the transfer function in terms of the 2×2 block matrices $\underline{\underline{\mathbf{A}}}, \underline{\underline{\mathbf{B}}}, \underline{\underline{\mathbf{D}}}, \underline{\underline{\mathbf{E}}}, \underline{\underline{\boldsymbol{\alpha}}}, \underline{\underline{\boldsymbol{\beta}}}, \underline{\underline{\boldsymbol{\gamma}}}, \underline{\underline{\boldsymbol{\zeta}}},$ and $\underline{\underline{\boldsymbol{\sigma}}}$.

Applying the boundary condition described above leads to the following expression

$$\underline{\underline{\mathbf{T}}}_F(\omega, I_0) =$$
$$\underline{\underline{\mathbf{Z}}}_L \left[ \left( \underline{\underline{\mathbf{A}}}\underline{\underline{\mathbf{Z}}}_L + \underline{\underline{\mathbf{B}}} + \underline{\underline{\mathbf{Z}}}_g \underline{\underline{\mathbf{D}}}\underline{\underline{\mathbf{Z}}}_L + \underline{\underline{\mathbf{Z}}}_g \underline{\underline{\mathbf{D}}} \right) - \left( \underline{\underline{\mathbf{Z}}}_g \underline{\underline{\boldsymbol{\beta}}} + \underline{\underline{\boldsymbol{\alpha}}} \right) \underline{\underline{\boldsymbol{\sigma}}}^{-1} \left( \underline{\underline{\boldsymbol{\gamma}}}\underline{\underline{\mathbf{Z}}}_L + \underline{\underline{\boldsymbol{\xi}}} \right) \right]^{-1}. \quad (B2)$$

In the calculation, the generator voltage is applied only at $z = 0$ in the first TL; with generator and load impedances of $\underline{\underline{\mathbf{Z}}}_L = \mathrm{diag}(Z_L, 0)$, and $= \underline{\underline{\mathbf{Z}}}_g = \mathrm{diag}(Z_g, 0)$, respectively, and in our numerical computations we take $Z_L = Z_g = 50\ \Omega$. Other combinations of generators and terminations setups are not considered here for brevity but they are rather important when optimizing the structure. Therefore, when computing (B2) we are only concerned with the 1,1-entry of the transfer function matrix $\underline{\underline{\mathbf{T}}}_F(\omega, I_0)$, which is a scalar function denoted simply by $T_F(\omega, I_0)$. In general the poles of the transfer function coincide with the system's natural frequencies [32], unless a pole-zero cancellation occurs that is however not encountered in the setup in Fig. 3. We numerically evaluate the complex frequency poles $\omega_p = \mathrm{Re}(\omega_p) + j\mathrm{Im}(\omega_p)$ of $T_F(\omega, I_0)$. We further check the multiplicity of the poles by applying Cauchy argument principle [31]. By choosing a contour C in the complex $\omega$ plane as shown in Fig. B1 we have

$$\oint_C \frac{T'_F(\omega, I_0)}{T_F(\omega, I_0)} ds = 2\pi j (N_z - N_p), \quad (B3)$$

where the prime symbol $T'_F(\omega, I_0)$ denotes the first derivative with respect to angular frequency, and C is a contour taken anticlockwise in the complex frequency plane such that $T_F(\omega)$ is analytic on C, and meromorphic inside C meaning it has isolated number of poles and zeros counted with their possible multiplicities [31], [32]. We apply the integral in (B3) in the vicinity of DBE real frequency $\mathrm{Re}(\omega) \sim \omega_d$ to check the multiplicity of the DBE pole in which (B3) gives numerically a value of $-2\pi j$ (with a relative error less than 0.1% due to numerical precision).